# Examining real-time TDDFT non-equilibrium simulations for the calculation of electronic stopping power

Dillon C. Yost, Yi Yao, and Yosuke Kanai*
*Department of Chemistry, University of North Carolina at Chapel Hill, Chapel Hill, NC 27516, USA*
(Dated: May 3, 2018)

In ion irradiation processes, electronic stopping power describes the energy transfer rate from the irradiating ion to the target material's electrons. Due to the scarcity and significant uncertainties in experimental electronic stopping power data for materials beyond simple solids, there has been growing interest in the use of first-principles theory for calculating electronic stopping power. In recent years, advances in high-performance computing have opened the door to fully first-principles non-equilibrium simulations based on real-time time-dependent density functional theory (RT-TDDFT). While it has been demonstrated that the RT-TDDFT approach is capable of predicting electronic stopping power for a wide range of condensed matter systems, there has yet to be an exhaustive examination of the physical and numerical approximations involved and their effects on the calculated stopping power. We discuss the results of such a study for crystalline silicon with protons as irradiating ions. We examine the influences of key approximations in RT-TDDFT non-equilibrium simulations on the calculated electronic stopping power, including approximations related to basis sets, finite size effects, exchange-correlation approximation, pseudopotentials, and more. Finally, we propose a simple and efficient correction scheme to account for the contribution from core electron excitations to the stopping power, as it was found to be significant for large proton velocities.

## I. INTRODUCTION

When an irradiating ion penetrates condensed matter, the ion transfers its kinetic energy via collisions with the nuclei and electrons in the target material. The stopping power for an ion penetrating a material is the key quantity used to describe this energy transfer [1] Stopping power is defined as the kinetic energy loss per unit distance of projectile ion displacement. For irradiating ions with high kinetic energies (> ~10 keV/nucleon), collisions with nuclei in the material are negligible. Instead, the ion's kinetic energy is transferred almost entirely via inelastic collisions to the electrons in the material. This phenomenon is known as electronic stopping, and its associated stopping power is referred to as electronic stopping power.

Electronic stopping power has been the focus of scientific research for decades due to its relevance in technological areas as wide-ranging as nuclear power [4,5], ion beam cancer therapy [7,8], aerospace materials [9,10], 3D ion beam lithography [11], and more. While quantifying electronic stopping power has become increasingly important in technology, experimental studies to determine electronic stopping power remain costly and complicated due to the need for advanced particle accelerators. At the same time, ever since the notion of electronic stopping was conceived, a wide range of analytical and simulation methods have been developed in an attempt to predict electronic stopping power and to provide deeper understanding into the mechanisms involved. Early approximated analytical models based on classical Coulomb scattering [12-14] were followed later by Bethe's quantum-mechanical perturbation approach [15] and Lindhard's calculations within the dielectric formalism based on the free-electron gas [16,17]. The formulae developed by Bethe and Lindhard both fall within the linear response formalism, and these approaches, with refinements and added corrections, remain some of the most widely used methods today. In the last 20 years, first-principles electronic structure theory has become available for calculating the microscopic dielectric response matrix in the context of linear response theory [18]. More recent work in this area has employed time-dependent density functional theory (TDDFT) formalism [19] in calculating the dielectric matrix, going beyond the random phase approximation in this context.

In more recent years, advancements in high-performance computing have opened up the possibility of studying electronic stopping by simulating the non-equilibrium response of electrons using first-principles simulations based on TDDFT. The great potential of this approach was demonstrated in work by Pruneda et al. [20] and Hatcher et al. [21]. Using the real-time propagation approach to TDDFT (RT-TDDFT) simulations [22], in which time-dependent Kohn-Sham equations are integrated in time, the non-equilibrium response of the electronic system to the projectile ion can be simulated. In the last 10 years, various aspects of the RT-TDDFT non-equilibrium simulation approach have advanced, and the approach has become increasingly popular and successful in studying electronic stopping in metals [23-27], semiconductors [6,28,29], and insulators [27,30], for a variety of projectile ions and a variety of condensed matter systems, including liquids

[30] and two-dimensional materials [31-34]. At the same time, there have been some indications that various physical and numerical approximations in these RT-TDDFT simulations could influence the calculated electronic stopping power curves. However, accurate experimental measurements of stopping power are rare for most materials, and discrepancies between RT-TDDFT and experimental results are hard to quantify.

Within the framework of linear response theory, where TDDFT (i.e. TD-LDA) was used to obtained the dielectric matrix, recent work by Shukri et al. [6,35] discussed that obtaining a converged result is challenging and requires extrapolation schemes, even for relatively simple materials like silicon. In light of these observations and the recent work by Shukri et al. [6,35], we reexamine the RT-TDDFT approach for calculating the electronic stopping power by considering crystalline silicon with a proton as the projectile ion, for which ample experimental results exist. In this article, we first briefly describe the computational methodology by which electronic stopping power is calculated from RT-TDDFT simulations. We then discuss the effects of a range of numerical and physical approximations on calculating the stopping power, identifying the prominent role of the pseudopotential approximation, which not only removes core electrons in calculations but also modifies the behavior of valence electron wavefunctions near the nuclei. We then describe and demonstrate a computationally feasible correction scheme based on single-atom collision, all-electron, RT-TDDFT simulations.

## II. METHOD

### A. Calculating Electronic Stopping Power from RT-TDDFT

We use the real-time propagation approach of TDDFT (RT-TDDFT) for obtaining electronic stopping power. In all simulations, the time-dependent Kohn-Sham (TDKS) equations are integrated for a system in which a classical, energetic proton with constant velocity travels through the material of interest (diamond cubic silicon), giving rise to a time-dependent external potential acting on the electronic system. A detailed description of the particular RT-TDDFT implementation used in this study can be found in Ref. [23].The average electronic stopping power can be obtained from the results of an RT-TDDFT simulation in a few different ways in practice. Electronic stopping power is defined as the rate of energy transfer from the projectile ion to the electronic system per unit distance of projectile travel. Thus, the most intuitive approach to calculate stopping power is to determine the slope of total electronic energy versus time given from the RT-TDDFT simulations. This approach requires a choice of linear regression and/or spline fitting, and it can influence the calculated stopping power, especially for higher projectile ion velocities. Alternatively, it can be shown that the non-adiabatic (NA) force on the projectile is equivalent to the instantaneous electronic stopping power [24]. Thus, for crystalline systems like diamond cubic silicon, we can average the NA force on the projectile over a distance that is an integer multiple of the periodicity of the crystal to precisely calculate the electronic stopping power for a given projectile path and velocity (see Supplemental Material at [URL] for details of the calculation of stopping power from NA force). In this work, we use the latter approach based on the NA force because of its unambiguity for crystalline systems like silicon.

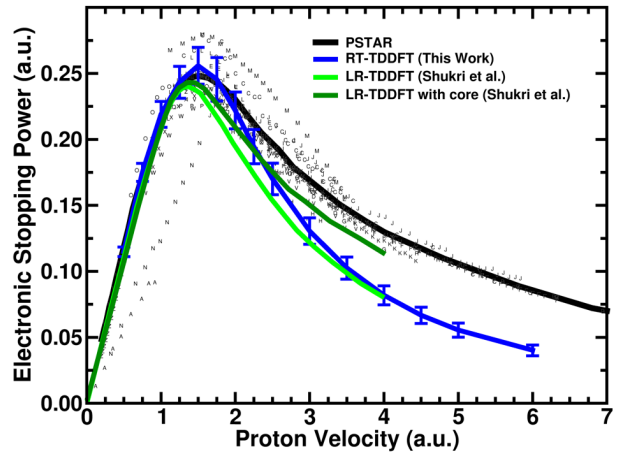

FIG. 1. Electronic stopping power as a function of the velocity of a proton projectile in silicon. The solid black curve represents the empirical PSTAR model [2]. The sets of letters correspond to different experimental data sets that can be found in Ref. [3]. The green curves represent results from the dielectric response formalism using LR-TDDFT for calculating the dielectric matrix [6] with the core electron correction (dark green) and without (light green). The solid blue curve corresponds to the values we obtained by calculating using RT-TDDFT with the "standard" parameters and approximations described in the Methods section. The error bars represent standard deviations of the mean stopping power along the given proton trajectory, which is calculated as an average of the instantaneous force on the projectile proton.

### B. Computational details

We first discuss the observed problem of underestimating the electronic stopping power at high proton velocities via the RT-TDDFT simulations in which the ground–state DFT calculations are sufficiently converged. In our plane-wave-pseudopotential (PW-PP) RT-TDDFT implementation in the Qb@ll/Qbox code [36], the time-dependent Kohn-Sham (TDKS) equations are propagated as the proton penetrates through silicon. We consider the case of the diamond crystalline phase of silicon because there exists a large amount of experimental data. Also, various approximations, such as the semi-local exchange-correlation (XC) potential, in DFT are sufficiently accurate for ground-state properties of silicon. We use the Perdew-Burke-Ernzerhof (PBE) XC functional [37], and we use the adiabatic approximation for its time-dependence [38,39].

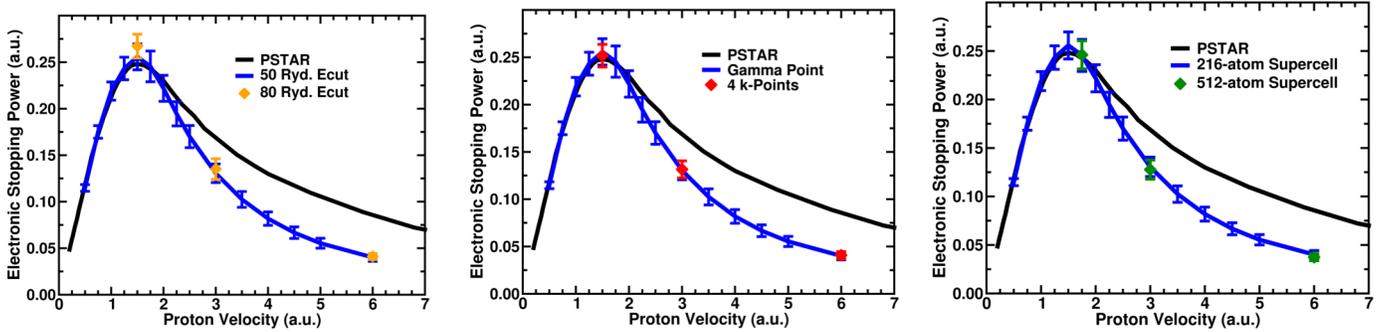

FIG. 2. Electronic stopping power as a function of the velocity of a proton projectile in silicon. The black curve represents the empirically fitted PSTAR model [2]. Convergence of the electronic stopping power curves with respect to various parameters are shown: (left) Calculations with 80 Rydberg planewave kinetic cutoff energy and 50 Rydberg planewave kinetic cutoff energy. (center) Calculations with 4 k-points in the Brillouin zone and only the gamma point in the Brillouin zone. (right) Calculations with at 512-atom supercell and a 216-atom supercell.

Norm-conserving pseudopotentials [40] of the Hamann-Schluter-Chiang-Vanderbilt type [41] were used for all atoms, including the projectile proton. Within this pseudopotential (PP) framework, the 4 valence electrons of each silicon target atom were treated explicitly. The PP cutoff radii of the 3s and 3p electrons of Silicon are 1.1 a.u. and 1.3 a.u., respectively. The PP cutoff radius for the 1s orbital of the Hydrogen projectile atom is 1.0 a.u. We used a large cubic simulation supercell of 216 atoms with the experimental lattice constant of 10.26 Bohr. Prior to performing the RT-TDDFT simulations, the ground state DFT calculation on the crystalline silicon supercell is carried out using the aforementioned computational parameters. It should be noted that no projectile ion is included in this ground-state DFT calculation because the purpose here is to acquire the unperturbed silicon electronic density to be used as the initial condition for the RT-TDDFT simulations. In the RT-TDDFT simulations, the projectile ion, a proton, is treated classically on an equal footing with all of the other ions in the simulation cell, with the exception that it moves at a constant velocity through the simulation cell, all other atoms being held fixed. We used only the gamma point for Brillouin zone sampling. The planewave (PW) cutoff is 50 Ryd.

This "standard" computational procedure is based on parameters that are sufficiently converged at the ground state DFT level for most ground-state properties, including the band gap (0.60 eV). The TDKS equations are integrated using an enforced time reversal symmetry (ETRS) propagator [42], with an integration time step of 0.5 attoseconds. The convergence for the stopping power calculation was confirmed by comparing to a smaller time step of 0.25 attoseconds (see Supplemental Material at [URL] for comparison of stopping power calculated with different time steps).

Instead of converging a classical ensemble average of projectile paths, the so-called "centroid path approximation" for the projectile path as demonstrated for 2D materials in [33] and validated for 3D materials in [28] was used in order to reduce the very large computational cost of RT-TDDFT simulations. As shown in our earlier work for the electronic stopping power of silicon carbide [28], the centroid path approximation does not introduce noticeable errors: The stopping power curves agree well between the stopping power calculations with and without the approximation for the computational framework described above.

In Figure 1, the stopping power results from the RT-TDDFT simulations are shown in comparison to the PSTAR empirical model [2], a wide array of experimentally reported values from the IAEA Nuclear Data Services database [3], and the LR-TDDFT results by Shukri, et al [6]. Exact details of each experimental data set can be found in Ref. [3] and the references therein. This figure and all of the following figures are presented in atomic units. The calculated RT-TDDFT values are in good agreement with the PSTAR model in the velocity regime at and below the Bragg peak (v ≤ ~1.5 a.u.). However, continuing into the high velocity regime, the RT-TDDFT simulation results significantly underestimate the electronic stopping power (52.5% difference at v = 6.0 a.u.) We can be confident that this underestimation does not indicate inaccuracy of the PSTAR model since all reported experimental data agree among them, showing a higher stopping power than our RT-TDDFT result. We note here that this underestimation is not unique to the target material of crystalline silicon, nor is it unique to the proton projectile ion. Previous RT-TDDFT studies indeed have shown similar degrees of disagreement with experiment for electronic stopping power for high-velocity protons and α particles in water [30], aluminum [24], and silicon carbide [28]. We also note that the relativistic Bethe formula [43] shows that taking the relativistic effect into account changes the stopping power only by +0.005% at v = 6.0 a.u. since the proton velocity is still only a small fraction of the speed of light even at such a large velocity. In Figure 1, we also include the corresponding stopping power curve from the linear response theory calculations in which the dielectric matrix was obtained from TDDFT [6]. While this result by Shukri et al. also showed the underestimation at high velocities, accounting for the semi-core electrons (2s and 2p electrons of silicon atoms) using an extrapolation scheme largely corrected the underestimation. Motivated by this work using linear response theory, we systematically revisit our non-equilibrium dynamics simulation approach based on RT-TDDFT and examine computational parameters and physical approximations in the following sections.

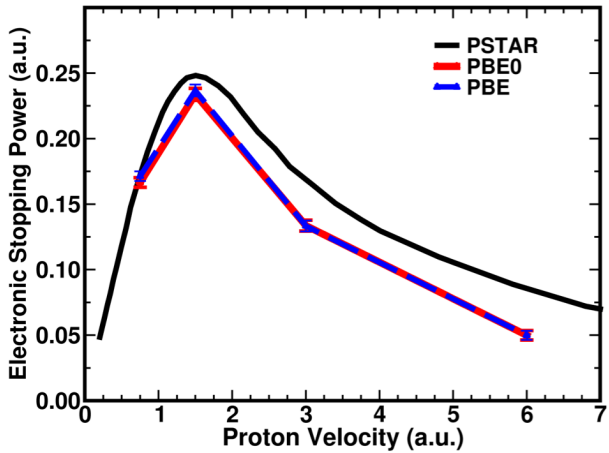

FIG. 3. Electronic stopping power as a function of the velocity of a proton projectile in Si. The solid black curve represents the empirical PSTAR model [2]. The colored curves correspond to the values we obtained by calculati using mixed Gaussian planewave (GPW) RT-TDDFT using the PBE XC functional (dashed blue) and the PBE0 hybrid XC functional (red). The error bars represent standard deviations of the mean stopping power along the given proton trajectory, which is calculated as an average of the instantaneous force on the projectile proton.

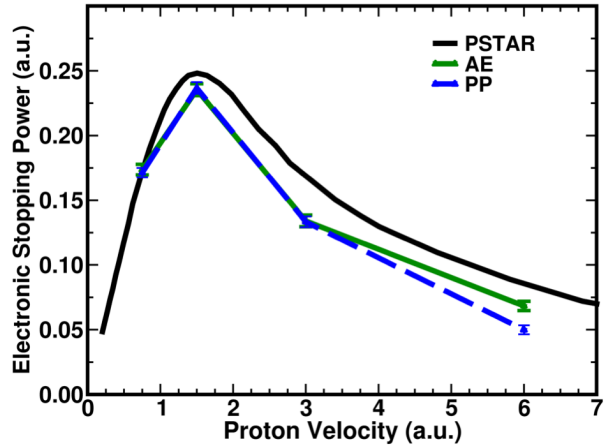

FIG. 4. Electronic stopping power as a function of the velocity of a proton projectile in Si. The solid black curve represents the empirical PSTAR model [2]. The colored curves correspond to the values we obtained by calculating using mixed Gaussian planewave (GPW) RT-TDDFT in which core electrons are treated using pseudopotentials (blue) and treated explicitly using the all-electron approach (green). The error bars represent standard deviations of the mean stopping power along the given proton trajectory, which is calculated as an average of the instantaneous force on the projectile proton.

## IV. RESULTS AND DISCUSSION

### A. Convergence of computational parameters

#### 1. Basis set

In order to examine the influence of the basis set in the plane-wave expansion, we used a higher plane-wave (PW) kinetic energy cutoff of 80 Rydberg, which is much larger than that which is required for converging ground-state properties, including the band structure. Due to the increased computational cost, we examined stopping power for only three velocities: $v$ = 6.0 a.u. (high velocity), $v$ = 3.0 a.u. (mid-range velocity) and v = 1.5 a.u. (near the Bragg peak). As can be seen in Figure 2, the 80 Rydberg simulation result shows no significant difference from the 50 Rydberg simulation (+2.5%) for v = 6.0 a.u. For v = 1.5 a.u., there is a slightly larger difference (+4.5%). The underestimation of electronic stopping power at high velocities cannot be attributed to incompleteness of the basis size.

#### 2. Brillouin zone sampling

Another possible contributor to the stopping power high velocity underestimation problem is insufficient sampling of the Brillouin zone with the single Gamma point approximation. We performed RT-TDDFT simulations for three proton velocities (v = 1.5 a.u., v = 3.0 a.u., and v = 6.0 a.u.) using 4 Monkhorst-Pack k-points. As can be seen in Figure 2, increasing the k-point sampling in the Brillouin zone has no significant effect on the calculated electronic stopping power values. Considering that a large supercell containing 216 silicon atoms is used, it is not surprising that the gamma point is sufficient to sample the Brillouin zone.

#### 3. Finite-size effect

Finite size errors could possibly be responsible for the stopping power underestimation at high velocities. While the convergence of the calculation with respect to the k-point sampling in the Brillouin zone is not a problem as discussed above, non-physical interactions among periodic images need to be examined in the calculations with the periodic boundary conditions (PBC). Note that a constant shift in the total energy as in the Makov-Payne correction [44] does not change the stopping power values. Figure 2 shows the stopping power using a large 512-atom supercell with 2048 electrons. As can be seen, finite size errors are negligible, and they do not explain the underestimation of our electronic stopping power curve for the high velocity region.

## 4. Exchange-correlation approximation

In the "standard" simulations, we employed the non-empirical generalized gradient approximation (GGA) functional, PBE [37], for our studies because of its balanced accuracy and efficiency, and the GGA XC approximation is particularly convenient for planewave (PW) implementations. At the same time, there exist other more advanced XC approximations such as PBE0 [45]. However, the computational cost of hybrid XC functionals like PBE0 is prohibitively large in PW-PP implementations because of the calculation of the Hartree-Fock (HF) integral. The majority of approximations/techniques that are used in the PW-PP framework for hybrid XC functionals do not translate well to an RT-TDDFT implementation. For this reason, we used the RT-TDDFT implementation in the CP2K code based on the mixed Gaussian and planewave approach (GPW) [46] to examine the influence of the XC approximation. In this approach, KS wavefunctions are represented using Gaussian functions, while the electron density is represented using plane-waves (i.e. GPW/GAPW formalism). We used the ETRS integrator with a time step of 0.5 attoseconds. The DZV3P Gaussian basis set was used for the KS wavefunctions, with Goedecker-Teter-Hutter (GTH) pseudopotentials for core electrons [47]. The electron density is represented in planewaves with an energy cutoff of 250 Ryd. The HF exchange in PBE0 was computed using the auxiliary density matrix method with the cFIT3 auxiliary basis set [48]. Our test with different basis sets show that DZV3P basis set is sufficiently large for describing the electronic stopping for this material (see Supplemental Material at [URL] for basis set convergence data).

We compare the RT-TDDFT electronic stopping power using the PBE and PBE0 functionals as implemented in the Libxc library [49]. Because of the high computational cost of the PBE0 approximation, we performed simulations only for a few selected proton velocities of interest: v=0.5, 1.5, 3.0, and 6.0 a.u. Figure 3 shows no significant difference between the PBE and the PBE0 approximations, especially at the high velocities. While the electronic stopping power at these velocities show no significant difference between the PBE and PBE0 XC functionals, the same may not be true for the threshold velocity at which the stopping power diminishes at low kinetic energies: This is due to the fact that the threshold velocity is directly related to the energy gap of the material [29], which can be quite sensitive to the choice of XC approximation.

## 5. Pseudopotential approximation

Using linear-response TDDFT and the random-phase approximation (RPA) to calculate the dielectric matrix within the framework of Lindhard's linear response model, Shukri, et al. reported the electronic stopping for silicon [6]. The linear response approach is generally considered to be reliable for large velocities. However, Shukri, et al. showed that the stopping power was significantly underestimated at large velocities, similar to our predictions based on RT-TDDFT. In this case of underestimation, however, semi-core electrons were not taken into account when calculating the electronic stopping power (see Figure 1). In the same work, Shukri, et al. proposed an extrapolation scheme to obtain the stopping power curve because converging numerical parameters such as the number of empty bands, number of k-points, and cutoff energy is currently challenging, especially for semi-core electrons (2s and 2p electrons of silicon atoms). Using the extrapolation scheme, and incorporating semi-core electrons, they showed that the electronic stopping power increased by 41% at v = 4.0 a.u. (see Figure 1). They did not consider the contribution to the stopping power from 1s electron in silicon atoms.

Within our "standard" RT-TDDFT simulations in the PW-PP framework, there are two notable deficiencies. One deficiency is that core electrons (i.e. 1s, 2s, and 2p electrons of silicon atoms) are not treated explicitly in the simulations, and therefore the calculated stopping power does not take into account any electronic excitations from core electrons. Second, using norm-conserving PPs, the wavefunctions for valence electrons do not behave properly within the PP cutoff radius. The total volume taken up by the PP volume amounts only to 5% ($r_{cut}$=1.1 a.u.) ~ 7% ($r_{cut}$=1.3 a.u.) of the total simulation cell volume, but the excitations from the core electrons likely become more dominant in electronic stopping at large velocities and during close ion-atom collisions. Unfortunately, it is prohibitively computationally expensive to represent all core electrons using plane-waves due to the fact that cutoff energies of hundreds of Rydbergs are required for convergence. Thus, we again used the RT-TDDFT simulations based on the Gaussian and augmented plane-wave method (GAPW) as implemented in the CP2K code for all-electron (AE) calculations to examine this aspect [46,47]. For AE calculations, the 6-311G**3P basis set was used, in which the 6-311G** basis set [50] was combined with 3 additional polarization basis functions taken

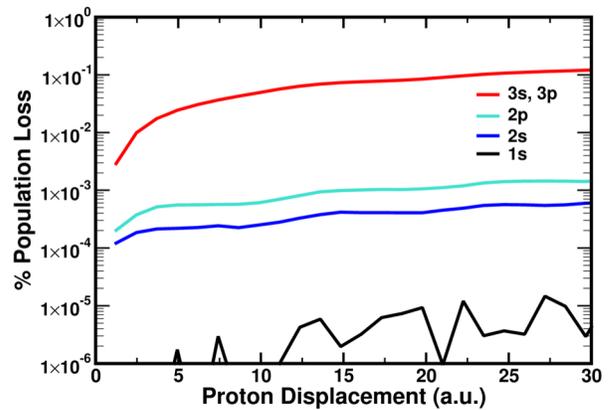

FIG. 5. Population analysis based on projections of the time-dependent Kohn Sham (TDKS) states onto the ground eigenstates of the silicon system. Percentages of population loss relative to the initial, fully occupied condition for the 3s and 3p atomic orbitals (red), the 2p orbitals (light blue), the 2s atomic orbitals (blue), and the 1s atomic orbitals (black) of the silicon atoms.

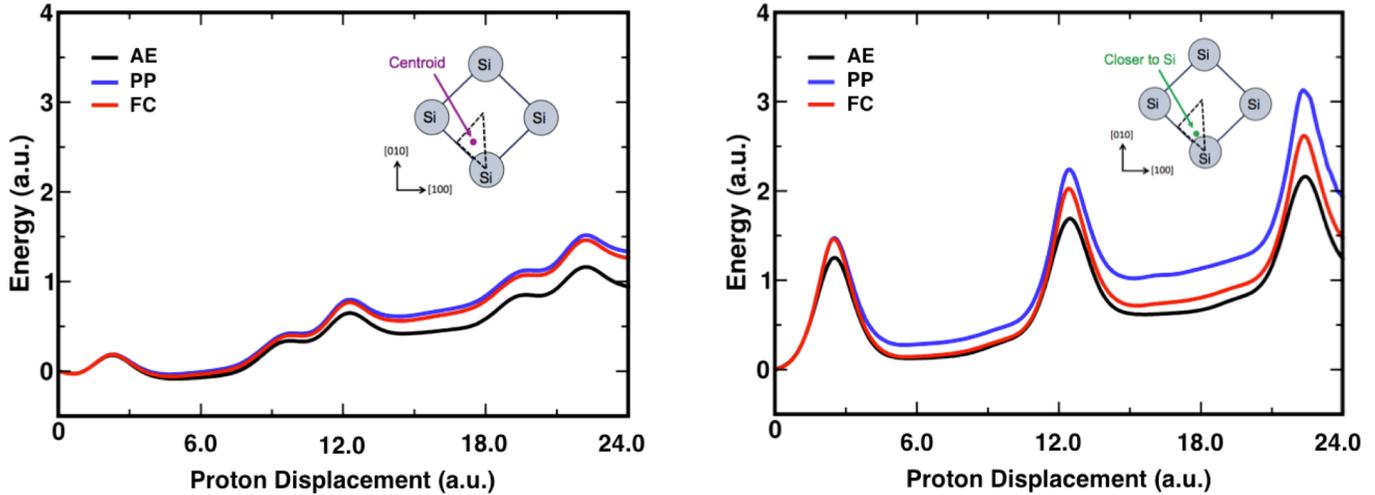

FIG. 6. Total energy as a function of projectile proton (v=6.0 a.u.) displacement in the simulation cell acquired from the RT-TDDFT simulations using the mixed GPW/GAPW approaches. In each plot, the results from three different treatments for the core (1s, 2s, 2p) electrons are shown: The black curve represents the simulation results in which the core electrons are represented by pseudopotentials (PP). The blue curve represents the all-electron (AE) results from using the GAPW method in which the core electrons are treated explicitly and allowed to be excited. The red curve represents the frozen-core (FC) results in which the AE wavefunctions are used but the core electrons are not propagated in time. (left) The results from the simulations for the centroid path trajectory. (right) The results from the simulations for a path parallel to the centroid path trajectory but 33% closer to the channel of target Si atoms.

from the DZV3P basis sets. Our test shows that DZV3P and the 6-311 G**3P basis sets yield the same stopping power (see Supplemental Material at [URL] for basis set convergence data). With the AE simulations, the integration step size needed to be reduced by approximately 50% for numerical stability for the time integration. F

Figure 4 shows the comparison between the all-electron (AE) simulation results and the pseudopotential (PP) simulation results. At the high velocity of v = 6.0 a.u., there is a noticeable difference between the AE and PP simulation results. The stopping power is approximately 30% higher with the AE calculation compared to the PP calculation. In order to analyze the contribution from core electrons to the underlying electronic excitations, we projected the time-dependent Kohn-Sham (TDKS) wavefunctions onto the eigenstates of the system (see Supplemental Material at [URL] for more information on the TDKS projections results). The eigenstates that derived from 1s, 2s, and 2p orbitals of silicon atoms are at –1776 eV, -134 eV, -91 eV below the valence band maximum (VBM), which is mainly composed of the 3s and 3p orbitals of silicon atoms. The projection of the TDKS wavefunctions onto these eigenfunctions shows that the occupations of the valence band (VB) eigenstates decrease due to the electronic excitations that take place in the electronic stopping process, as expected. The time-dependent changes of the occupation for the eigenstates for the 1s, 2s, 2p, and valence electrons at the proton velocity of 6 a.u. are represented in Figure 5. For the 1s state, the occupation decreases only slightly. For the 2s, and 2p states, the decreasing occupations are more substantial. Using the KS eigen-energies and the changes in the occupations, we estimate the extent to which the electronic stopping power is due to the single-particle excitations of these core electrons. The relative contributions of 1s, 2s, and 2p excitations to the stopping power are estimated to be approximately 0.3%, 1.5%, and 7.9% of the contribution from valence electron excitations. The result clearly shows that the PP approximation can be problematic in simulations of electronic stopping at high velocities. Importantly, this finding also calls for a reexamination of the validity of the centroid path approximation [28,33] for the large velocities because along this path, the proton projectile does not come near the target silicon atoms.

### B. Analysis of core electron effects

In addition to the AE and PP calculations, we also performed AE simulations in which the core electrons are not propagated in time. In these "frozen core (FC)" simulations, core electrons are restricted but the valence-electron wavefunctions correctly behave near the nuclei (except at the cusp). Figure 6 shows the comparison among the all-electron (AE) calculation, the frozen-core (FC) calculation, and the pseudopotential (PP) calculation results. In each of these simulations, we used the centroid path approximation (CPA) in which the closest distance from the projectile proton to a silicon atom is 1.9 a.u. Fig 6 (a) shows that the FC result is much closer to the AE result than to the PP result. Thus, for the CPA trajectory, the difference between the AE and PP calculations does not arise mainly as a result of core electron excitations, but rather it is the valence-electron wavefunction behavior close to the nuclei that is responsible. Fig 6 (b) shows the same set of simulations, but now the proton projectile path is closer to silicon atoms such that the minimum projectile-target distance is ~1.3 a.u (instead of the 1.9 a.u. in the CPA). In this case, the AE-vs-PP difference primarily derives from core electron excitations. The analysis based on the KS eigen-energies shows that the relative contributions of 1s, 2s, and 2p excitations to the stopping power are approximately 0.6%, 6.6%, and 22.9% of the contribution from valence electron excitations, being much larger than the centroid path trajectory. Although the centroid path approximation has been found to be satisfactory for calculating the electronic stopping power of various materials using simulations with pseudopotentials, this approximation becomes less satisfactory when core electron effects are important as shown in the all-electron RT-TDDFT calculations. At high velocities, semi-core electron excitations contribute significantly to the electronic stopping. Unfortunately, converging the stopping power with respect to the ensemble average of proton trajectories in all-electron RT-TDDFT simulations (3024 electrons) comes at a prohibitive computational expense. This is due to two factors: Small regions of the total simulation cell (< 10%) dominate the excitations at high velocities, implying that a large number of random trajectories would be required to converge the ensemble average. Additionally, stable numerical integration of the TD-KS equations becomes highly difficult during close collisions between the projectile ion and target atom nuclei, requiring a very small discrete time step, and this is also coupled with the fact that very high planewave kinetic cutoff energy is required if the PW-PP formalism is used.

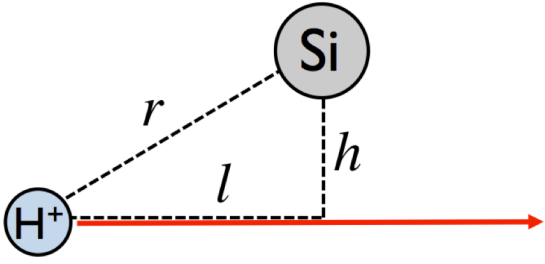

FIG. 7. Scheme for the single-atom collision simulations, where the red arrow indicates the trajectory of the proton (H$^+$). The simulation is carried out over a range of impact parameters $h$. The distance between the proton and the Silicon target atom is given by $r$. The displacement of the proton is given by $l$.

### C. Proposed correction scheme

Here we propose a simple and efficient correction scheme to account for the missing "core electron" effects. We consider obtaining the corrected electronic stopping power $\bar{S}(v)$ for silicon by scaling the stopping power that is obtained from the RT-TDDFT simulations by a velocity-dependent correction factor $F_{core}(v)$, that takes into account the core-electron effects,

$$\bar{S}(v) = S_{CP+PP}(v) \cdot (1 + F_{core}(v)) \quad (1)$$

where $S_{CP+PP}$ is the stopping power acquired in the "standard" simulations using the centroid path approximation and the PP approximation. The scaling factor is given in terms of the ratio between the core and valence electron contributions to the electronic stopping for a *single* silicon atom such that

$$F_{core}(v) = \sum_{i=1}^{N} \int dV \frac{S_{core}^{Atm}(v,r_i)}{S_{val}^{Atm}(v,r_i)} \quad (2)$$

where the integrand is the ratio between the core and valence electron contributions, which are functions of the distance between the projectile proton and the target silicon atom (in addition to the velocity dependence). The volume integral contains all of the projectile proton position in the simulation cell, and the variable $r_i$ is the distance to all $N$ silicon atoms in the simulation cell. In practice, it is not necessary to consider the atoms that are beyond a small radius (~5 a.u.) of the projectile ion since $S_{core}^{Atm}$ becomes zero at large distances. The distance-dependent $S_{val/core}^{Atm}$ are related to the total energy transfer that depends on the impact parameter, $h$, as (see Figure 8)

$$\int_{-\infty}^{\infty} S_{val/core}^{Atm}(v,r) dl = \Delta E_{val/core}(v,h) \quad (3)$$

We observed from the single-atom RT-TDDFT simulations that, for a given projectile ion velocity, the energy transfer function follow closely a Gaussian function

$$\Delta E_{val/core}(v,h) = A_{val/core}(v) e^{-B_{val/core}(v)h^2} \quad (4)$$

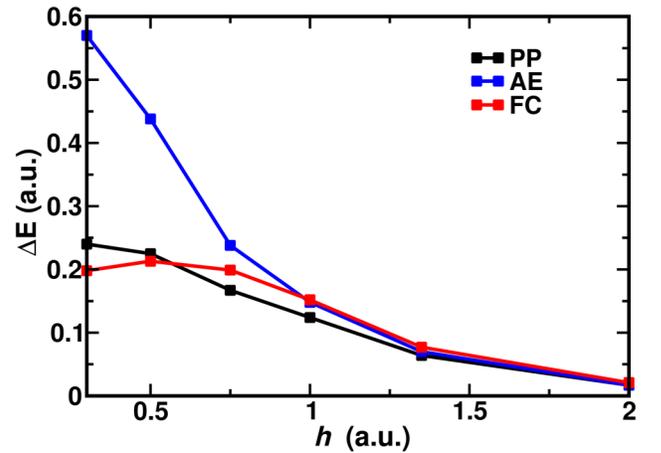

FIG. 8. Energy transferred in single-atom collision simulations as a function of impact parameter $h$ for the simulations using pseudopotentials (black), all electrons (blue), and frozen core (red) schemes. See the text for details.

By expressing $S_{val/core}^{Atm}$ also as

$$S_{val/core}^{Atm}(v,r) = C_{val/core}(v)e^{D_{val/core}(v)r^2} \quad (5)$$

for numerical convenience. Geometric relationships between $r$, $h$, and $l$ of Figure 7 then yield

$$D(v) = B(v) \quad (6)$$

and

$$C(v) = A(v)\sqrt{\frac{B(v)}{\pi}}. \quad (7)$$

We can thus determine these coefficients $D(v)$ and $C(v)$ via Eq. 6 and Eq. 7 by finding the parameters $A$ and $B$ in Eq. 4 by obtaining $\Delta E_{val/core}$ from "single-atom simulations" in which the impact parameter between the projectile proton and a single target silicon atom is varied. $\Delta E_{core}$ is obtained from the AE RT-TDDFT and with the PP approximation as

$$\Delta E_{core}(v,h) = \Delta E_{AE}(v,h) - \Delta E_{val}(v,h). \quad (8)$$

Figure 8 shows the energy transferred versus impact parameter for the single collision simulations for v=6.0 a.u. For impact parameters below ~0.75 a.u., the all-electron simulations show significantly higher energy transfer than the PP simulations. The results are fit to the Gaussian function (Eq. 4) to determine the parameters A and B. Having determined the shape of $S_{val/core}^{Atm}$ functions, the velocity-dependent scaling factor is determined to correct for the missing core effects as in Eq. 2.

This correction scheme can be also applied using PW simulations using semi-core PPs, in addition to the all-electron simulations with the GAPW approach. In the semicore PPs, only the 1s electrons are pseudized, and the 2s, 2p, 3s, and 3p electrons of silicon atoms are treated explicitly. In the expressions given above, the energy transfers in the semi-core PP simulations are used in place of the all-electron single-atom energy transfers. This alternative approach is possible due to the fact that 1s electron excitations contribute less than 1% to the total electronic stopping power. For these single-atom collision simulations, we generated semicore pseudopotentials for Silicon with cutoff radii of 0.40, 0.35, and 0.40 Bohr for the 2s, 2p, and 3p orbitals, respectively, in keeping with the parameters used by Tiago et al. [51]. Due to the very small cutoff radii, a very large planewave cutoff energy of 400 Rydberg was required to converge the energy gap. While this high cutoff energy would make RT-TDDFT simulations with the Silicon supercell prohibitively expensive, the computational cost is manageable for the single-atom collision simulations.

Figure 9 and Figure 10 show the results of applying the core correction factors for the GPW-based (corrected with all-electron single-atom calculations) simulation result and PW-based simulation result (corrected with semi-core PP single-atom calculations), respectively. When these corrections are applied, we observe significant improvement with respect to the experimental data. The corrected GPW and PW electronic stopping power curves show the average absolute % difference relative to the PSTAR results of 5.1% and 5.2%, respectively. Applying the core correction increases the stopping power by -55% (105%) at v = 6.0 a.u. for the GPW (PW) simulations. For v <= 1.0 a.u., there is negligible difference between the corrected and uncorrected results. At the Bragg peak (v = 2.0 a.u.), the core correction increases the GPW (PW) electronic stopping power by 9.0% (7.1%).

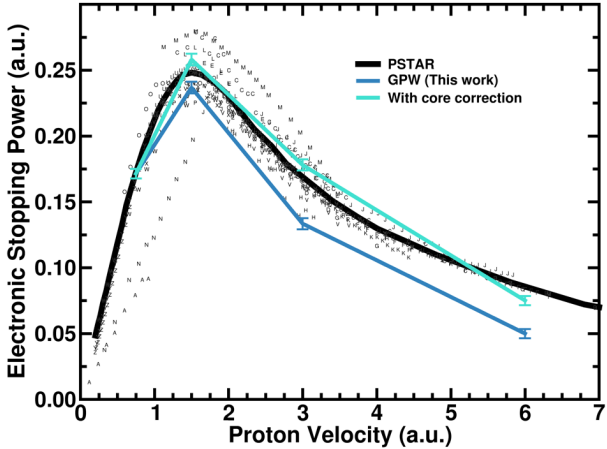

FIG. 9. Electronic stopping power as a function of the velocity of a proton projectile in Si. The solid black curve represents the empirical PSTAR model [2]. The sets of letters correspond to different experimental data sets that can be found in Ref. [3]. The dark blue curve corresponds to the values we obtained by calculating using mixed Gaussian planewave (GPW) RT-TDDFT with the "standard" parameters and approximations described in the Methods section. The light blue curve corresponds to the results after the core correction has been applied. The error bars represent standard deviations of the mean stopping power for the trajectory.

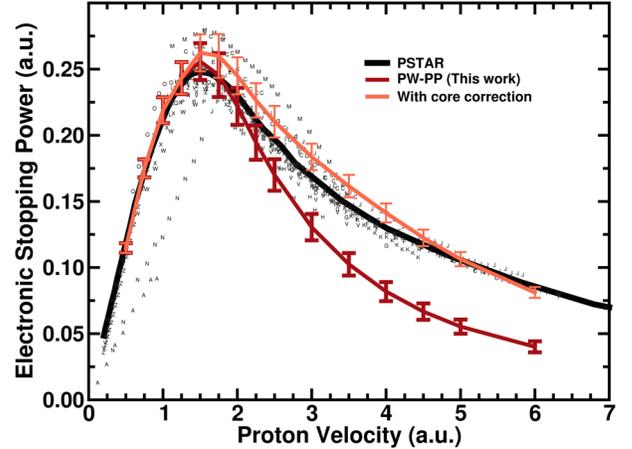

FIG. 10. Electronic stopping power as a function of the velocity of a proton projectile in Si. The solid black curve represents the empirical PSTAR model [2]. The sets of letters correspond to different experimental data sets that can be found in Ref. [3]. The dark red curve corresponds to the values we obtained by calculating using planewave pseudopotential (PW-PP) RT-TDDFT with the "standard" parameters and approximations described in the Methods section. The light red curve corresponds to the results after the core correction has been applied. The error bars represent standard deviations of the mean stopping power for the trajectory.

## V. CONCLUSIONS

We presented a thorough, methodical examination of non-equilibrium electron dynamics simulations based on RT-TDDFT for the calculation of electronic stopping power. We have shown that converging several computational factors, such as k-point sampling, planewave cutoff energy, and exchange-correlation (XC) approximation, with respect to the ground state DFT calculations is sufficient to ensure convergence of RT-TDDFT results. However, we have also identified several approximations that require special attention in these types of simulations. The combination of the centroid path approximation and the pseudopotential approximation leads to a neglect of core electronic effects, which are particularly important at high projectile ion velocities. We showed that core electrons affect electronic stopping in two ways: core electron excitations and modification of the valence wavefunctions near the core regions. We demonstrated that both of these effects contribute to an increase in electronic stopping and that the relative importance of these effects has a strong impact-parameter dependence. Unfortunately, performing RT-TDDFT simulations with all-electrons or semi-core-PPs setup for a classical ensemble of projectile ion trajectories is not yet practical due to the computational expense. As an alternative approach, we proposed a single-atom correction scheme to approximately take into account the missing core electronic excitations for the electronic stopping power calculation.

Another approximation in this work is the adiabatic approximation to the XC potential in RT-TDDFT. This approximation results in having the XC potential that depends only on the instantaneous electron density, neglecting any memory effects [52]. Nazarov *et al.* [53] have shown, within the linear response formalism, that error resulting from the adiabatic approximation is negligible for low-Z ions such as protons and α particles stopping in a homogeneous electron gas. However, for high-Z ions, the error can be significant. The influence of the adiabatic XC approximation in the RT-TDDFT approach on electronic stopping power needs to be investigated in the future. In the RT-TDDFT approach, having the energy functional as a constant of motion is essential for directly obtaining the electronic stopping power from the electronic energy [23,31], and the stopping power can be calculated equivalently from the Hellman-Feynman force on the projectile ion, as we have done here, when the adiabatic approximation is used. However, the XC functional, in principle, depends on the electron density at previous times and on the initial wavefunction. As discussed by Ullrich in the context of the current TDDFT [54], deriving an accurate XC approximation that ensures a constant total energy is quite complicated when the adiabatic XC approximation is alleviated. If this were accomplished, one could then also derive the corresponding expression for forces from the resulting action [55], which is unlikely to be the Hellman-Feynman force when the adiabatic XC approximation is not adapted. Going beyond the adiabatic approximation in RT-TDDFT for simulation of real materials is presently out of reach and remains an active area of current research [56,57].

## ACKNOWLEDGMENTS

We acknowledge computational resources from the Argonne Leadership Computing Facility, which is a Department of Energy (DOE) Office of Science User Facility supported under Contract No. DE-AC02-06CH11357. An award of computer time was provided by the Innovative and Novel Computational Impact on Theory and Experiment (INCITE) program. This work is supported by the National Science Foundation under Grants No. CHE-1565714 and No. DGE-1144081.

*ykanai@ad.unc.edu

**Supplementary material for**

**"Examining Real-time TDDFT Non-equilibrium Simulations for the Calculation of Electronic Stopping Power"**

**Dillon C. Yost, Yi Yao, and Yosuke Kanai**

**I. Calculation of electronic stopping power from non-adiabatic "drag" force.**
The electronic stopping power for a given proton velocity was calculated by averaging the non-adiabatic force on the proton in the direction of its motion [1] . The force was averaged over a distance commensurate with the periodicity of the crystal structure as illustrated in FIG. S1.

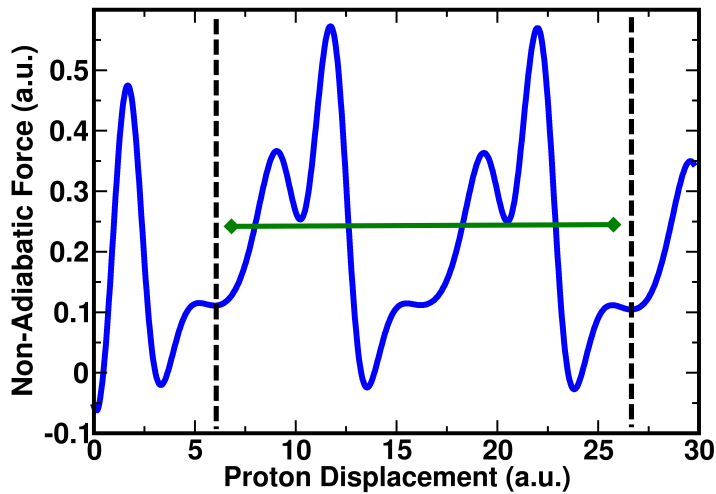

FIG. S1. The non-adiabatic drag force (blue curve) on the proton for v=2.0 a.u. The force is averaged over a distance of 20.526 Bohr (twice the lattice constant of Silicon), as indicated by the vertical dashed lines. The resulting electronic stopping power for a proton with v=2.0 a.u. is 0.25 a.u., as indicated by the horizontal green line.

**II. RT-TDDFT Numerical Integration Time Step Dependence**
As described in work by Schleife et al. [2], RT-TDDFT simulations can be sensitive to the discrete time step used in the numerical integrator. In this study, we used the enforced time-reversal symmetry (ETRS) integrator [3], which is typically numerically accurate even with time step as large as few attoseconds and even for rather high planewave cutoff energy. In order to test the convergence of our RT-TDDFT stopping power results with respect to the step size, we compared the standard simulation results (dt = 0.5 Attoseconds) to results acquired with a smaller time step of 0.25 attoseconds. As can be seen in Fig. S2., the RT-TDDFT simulations are well-converged with a time step of 0.5 attoseconds.

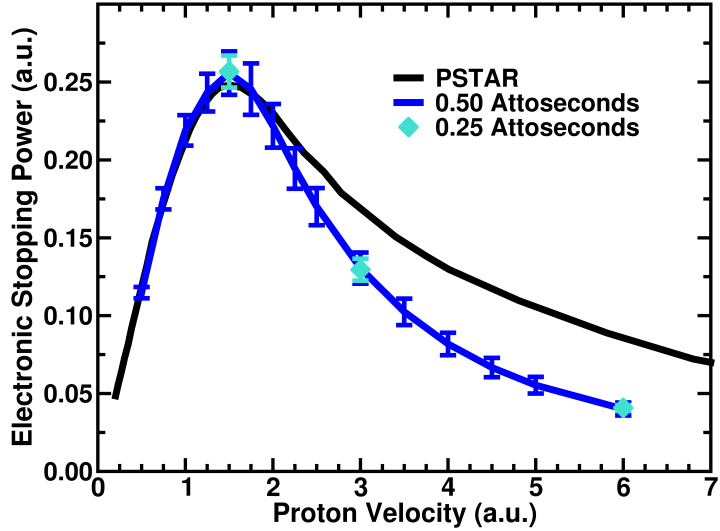

FIG. S2. Comparison of electronic stopping power calculated with RT-TDDFT using the ETRS propagator with two different time steps: 0.50 attoseconds (blue) and 0.25 attoseconds (light blue).

### III. Projections of TD-KS Wavefunctions onto Ground State Eigenfunctions

For the all-electron simulations, we calculated projections of the time-dependent Kohn-Sham (TD-KS) wavefunctions onto the eigenstates of the system at equilibrium. Using the KS single-particle eigenvalues and the time-dependent occupation changes in terms of the eigenstates of silicon at equilibrium, we can roughly estimate contributions from core (1s, 2s, 2p) electrons for the electronic excitations during the electronic stopping process. We considered both the centroid path and the path that is 33% closer to the channel of silicon atoms. The 1s state shows negligible contribution for both paths. The 2s and 2p states' contributions show a step-wise behavior that coincides with the periodicity of silicon atoms, and this step-wise behavior due to the silicon atoms is more evident in the close path case.

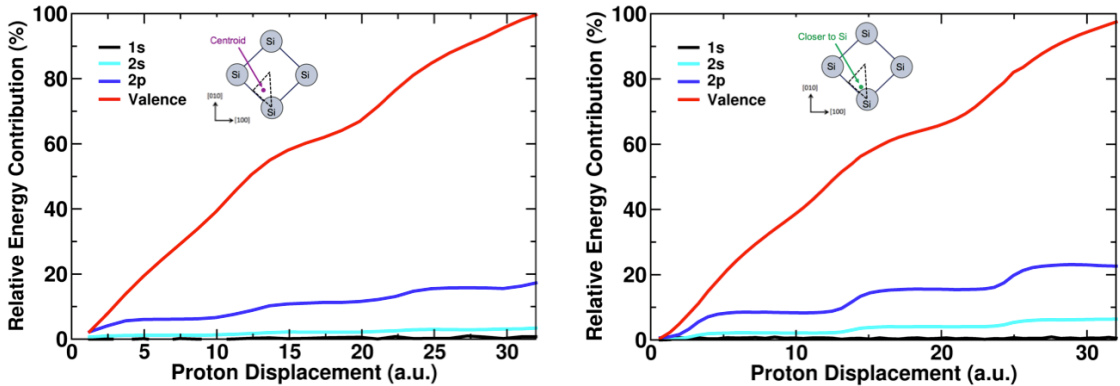

FIG. S3. Single-particle energy contributions from the 1s (black), 2s (light blue), 2p (blue), relative to that of valence (red) states in the all-electron RT-TDDFT simulations, for the centroid path (LEFT) and the path that is 33% closer to the target silicon atoms (RIGHT).

## IV. Effective Charge of the Projectile Proton

With RT-TDDFT simulations, one is provided with the time-evolving electron density of the system, information which allows one to calculate the effective charge state of the ion from first principles. However, in practice, this quantification of the charge state depends on a sensible charge partitioning scheme for the non-equilibrium electron density. We employed the Voronoi analysis using the code by Henkelman et al. [4]. 60 equally spaced "snapshots" of the electron density at different times were taken for each projectile velocity. Next, so-called "induced electron densities" were calculated by subtracting the cubic silicon ground state electron density from the non-equilibrium electron densities at the different times. These induced electron densities provide a spatial representation of where electron density is accumulating in the simulation cell and where it is being depleted. Finally, Voronoi analysis is carried out on the induced electron densities to quantify the charge within the projectile ion's Voronoi cell at different positions in the trajectory. While partitioning schemes based on electron density topology (e.g. Bader decomposition) are commonly used for the ground-state electronic density, the present problem lends itself better to the above approach using geometry-based Voronoi analysis [4] for the reasons outlined in Ref. [5]. The results are shown in FIG. S4.

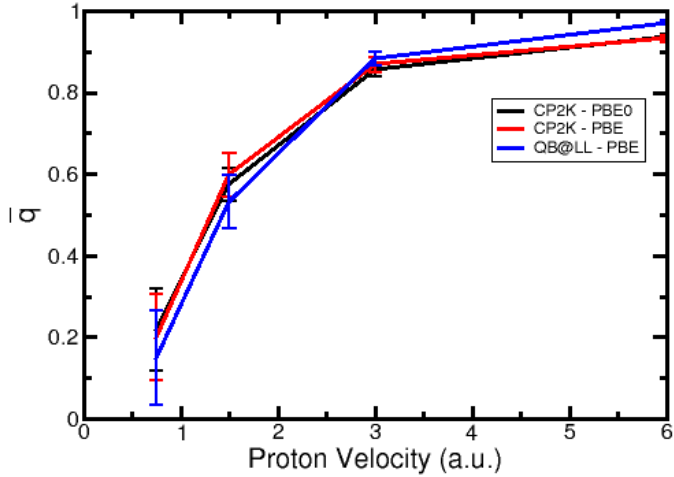

FIG. S4. Charge state of the projectile proton in RT-TDDFT simulations using both the planewave code (QB@LL) and the mixed Gaussian-planewave code (CP2K). The results from PBE and PBE0 simulations are shown for the CP2K code. All results follow the same qualitative trend: lower effective charge states at low velocities leading to almost full ionization at high velocities.

## V. Accelerated Voronoi charge analysis in real space

When performing Voronoi charge analysis to charge density information stored on a real space grid, each voxel in the volume needs to be assigned to an atom. A simplest way of doing so is looping over all the voxels, for each voxel, calculating its distances to all the atoms and finding the closest one. This approach yields an $O(n*m)$ computational cost, where n is the total number of voxels on the real space grid and m is the total number of atoms in the cell for analysis. For our work, we implemented an improved approach for the Voronoi analysis because a large number of charge density information needed to be analyzed. By combining a divide-and-conquer strategy and the so-called k-d tree algorithm[6], which is often used in computational geometry, we can reduce the computational cost to $O(\log(n)*\log(m))$ and yields a 100-500 times speed up for the real space charge density analysis. Divide-and-conquer strategy is, first, used to reduce the computational time of looping over all the voxels. For the divide-and-conquer method, we recursively divide the simulation cell into eight equivalent smaller cells, called octants. Then, for each octant cell, we check to see if there is a unique atom that is closest from all the eight corner voxels of the grid. If not, we continue to divide the octant further into eight equivalent smaller cells to create a new smaller octant until all octants have been assigned to an atom. This strategy cuts all unnecessary calculations for looping over the grid voxels in individual octant regions. This makes the computational cost of the looping over the real space grid to $O(\log(n))$ instead of $O(n)$. Furthermore, the k-d tree algorithm is used to accelerate searching of the closest atom from each grid voxel. The basic idea of the k-d tree is to perform a pre-process to all the atoms in the space and generate the k-d tree, and then using the k-d tree as an index-like tool. By using this algorithm, the search time for the closest atom is reduced to $O(\log(m))$[6]. We implemented these methods by modifying the standard Voronoi analysis implementation of the Bader analysis code by Henkelman, et al. [7]. The KDTREE2 code[6] is linked to handle the k-d tree algorithm. The performance of this new implementation (on a single processor) is shown for a bulk water system of 32 $H_2O$ water molecules with different real space grid sizes and also with a simulation cell containing 512-atom silicon and one proton cell.

| System | Real space grid | Time for original implementation (sec) | Time with divide-conquer method only (sec) | Time with divide-conquer method and k-d tree (sec) |
|---|---|---|---|---|
| 32$H_2O$ | 100^3 | 42.32 | 4.83 | 0.37 |
|  | 144^3 | 126.31 | 11.14 | 0.85 |
|  | 192^3 | 299.97 | 19.73 | 1.52 |
|  | 270^3 | 833.57 | 44.67 | 3.42 |
| 512 Si + 1 H | 192^3 | 1713.11 | 184.98 | 2.23 |

## VI. Dependence of RT-TDDFT calculation on Basis Set : CP2K code

Dependence on the Gaussian basis set were tested with SZV, DZVP, TZV2P, DZV3P, and 6-311G**3P basis sets for RT-TDDFT simulations with the proton velocity of 1.5 a.u. (at the Bragg peak) for the centroid path. FIG. S5 shows the electronic energy increase as a function of the projectile proton displacement. As can be seen in FIG. S5, the DZV3P, TZV2P, and 6-311G**3P basis sets show almost identical rates of energy increase. We used the DZV3P basis set for the simulations with pseudopotentials and the 6-311G**3P basis set for the all-electron simulations.

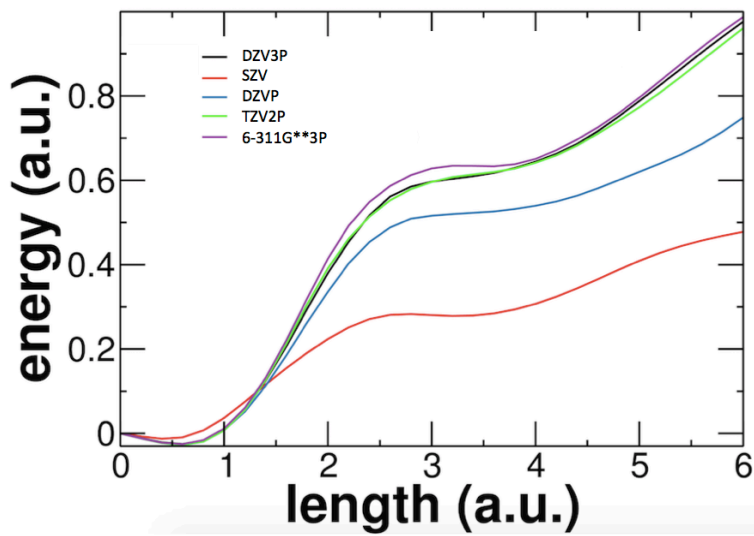

FIG. S5. Basis set dependence of the total energy change as a function of the projectile proton (v=1.5 a.u.) displacement in RT-TDDFT simulations using the CP2K code.

## Gaussian Basis sets used for RT-TDDFT calculations with the CP2K code

### For Pseudopotential calculations
```
Si DZV3P
 2
 3 0 1 4 2 2
      1.2032403600   0.3290356759   0.0000000000   0.0474536439   0.0000000000
      0.4688385970  -0.2533162616   0.0000000000  -0.2594495462   0.0000000000
      0.1679853910  -0.7870936517   0.0000000000  -0.5440932235   0.0000000000
      0.0575616890  -0.1909870193   1.0000000000  -0.3623984652   1.0000000000
 3 2 2 3 3
      0.8000000000   1.0000000000   0.0000000000   0.0000000000
      0.3000000000   0.0000000000   1.0000000000   0.0000000000
      0.1500000000   0.0000000000   0.0000000000   1.0000000000
```

### For All-electron calculations
```
Si 6-311G**3P
 11
 1 0 0 8 2
     69379.23000000         0.00075700         0.00000000
     10354.94000000         0.00593200         0.00000000
      2333.87960000         0.03108800         0.00000000
       657.14295000         0.12496700         0.00000000
       214.30113000         0.38689700         0.00000000
        77.62916800         0.55488800         0.17788100
        30.63080700         0.00000000         0.62776500
        12.80129500         0.00000000         0.24762300
 1 0 0 1 1
```

```
     3.92686600        1.00000000
 1 0 0 1 1
     1.45234300        1.00000000
 1 0 0 1 1
     0.25623400        1.00000000
 1 0 0 1 1
     0.09427900        1.00000000
 1 1 1 4 1
   335.48319000        0.00886600
    78.90036600        0.06829900
    24.98815000        0.29095800
     9.21971100        0.73211700
 1 1 1 2 1
     3.62114000        0.61987900
     1.45131000        0.43914800
 1 1 1 1 1
     0.50497700        1.00000000
 1 1 1 1 1
     0.18631700        1.00000000
 1 1 1 1 1
     0.06543200        1.00000000
 3 2 2 3 3
     0.8000000000   1.0000000000   0.0000000000   0.0000000000
     0.3000000000   0.0000000000   1.0000000000   0.0000000000
     0.1500000000   0.0000000000   0.0000000000   1.0000000000
```

## VII. Cartesian Coordinates of Silicon Atoms and Projectile Proton

The initial coordinates of the projectile proton (in atomic units) are (-15.3918, -1.7518, -0.8465). The unit vector of the proton's velocity is (1, 0, 0). The coordinates of the target silicon atoms in the 216-atom supercell are listed below:

```
Si -12.7702 -12.7678 -12.7714      Si -12.7643 7.75209 7.75587       Si -2.50983 7.74833 -2.50475
Si -12.7749 -7.63954 -7.64098      Si -12.7643 12.8831 12.8831       Si -2.50618 12.8807 2.62118
Si -7.63847 -12.7726 -7.6419       Si -7.6305 7.74597 12.8796        Si 2.6225 7.74608 2.61918
Si -7.64132 -7.63847 -12.7667      Si -7.63061 12.8819 7.75408       Si 2.62154 12.8784 -2.50817
Si -12.7726 -12.7714 -2.51265      Si -2.50758 -12.7726 -12.7691     Si -2.50243 7.74374 7.74857
Si -12.7784 -7.64179 2.61932       Si -2.51288 -7.63965 -7.63555     Si -2.50243 12.8784 12.8831
Si -7.64108 -12.7702 2.62154       Si 2.61966 -12.7714 -7.63683      Si 2.62435 7.74961 12.8831
Si -7.64766 -7.64283 -2.51017      Si 2.61449 -7.64098 -12.7691      Si 2.62414 12.8784 7.74925
Si -12.7691 -12.7691 7.75539       Si -2.50958 -12.7714 -2.50947     Si 7.74821 -12.7691 -12.7702
Si -12.7749 -7.63695 12.8843       Si -2.51536 -7.64272 2.61882      Si 7.74632 -7.63591 -7.64236
Si -7.6379 -12.7691 12.8854        Si 2.61896 -12.7714 2.61685       Si 12.8796 -12.7678 -7.64165
Si -7.64507 -7.64226 7.7528        Si 2.61862 -7.6379 -2.51065       Si 12.876 -7.63683 -12.7737
Si -12.7714 -2.50842 -12.7691      Si -2.50746 -12.7726 7.75173      Si 7.75162 -12.7702 -2.51111
Si -12.7702 2.62496 -7.63862       Si -2.51229 -7.64226 12.8843      Si 7.74891 -7.64132 2.618
Si -7.64497 -2.50664 -7.63483      Si 2.62036 -12.7737 12.8819       Si 12.8843 -12.7726 2.61779
Si -7.63965 2.62143 -12.7702       Si 2.6159 -7.64307 7.74807        Si 12.8772 -7.63731 -2.51322
Si -12.776 -2.50817 -2.50853       Si -2.51454 -2.50912 -12.7667     Si 7.74975 -12.7737 7.74704
Si -12.7749 2.62189 2.62317        Si -2.51372 2.62082 -7.63565      Si 7.74233 -7.63954 12.8784
Si -7.64955 -2.51464 2.62179       Si 2.61449 -2.50771 -7.64036      Si 12.8784 -12.7655 12.8807
Si -7.64507 2.62236 -2.50361       Si 2.61896 2.61954 -12.7702       Si 12.876 -7.6399 7.74843
Si -12.7784 -2.51171 7.75162       Si -2.51711 -2.51158 -2.50642     Si 7.74715 -2.5083 -12.7737
Si -12.7702 2.61896 12.8831        Si -2.51265 2.61296 2.62179       Si 7.7515 2.61954 -7.64118
Si -7.64297 -2.51171 12.8831       Si 2.61708 -2.51347 2.61918       Si 12.8807 -2.50654 -7.64236
Si -7.6419 2.61342 7.75009         Si 2.61661 2.61637 -2.50889       Si 12.8843 2.62118 -12.7726
Si -12.7643 7.75372 -12.7714       Si -2.51311 -2.51664 7.75043      Si 7.74833 -2.50654 -2.51183
Si -12.7678 12.8843 -7.64098       Si -2.50758 2.61473 12.8807       Si 7.75043 2.6205 2.62036
Si -7.63862 7.75268 -7.63695       Si 2.61519 -2.51183 12.8831       Si 12.876 -2.507 2.61918
Si -7.63437 12.8796 -12.7726       Si 2.6199 2.61509 7.74975         Si 12.8831 2.62379 -2.50947
Si -12.7691 7.75655 -2.5051        Si -2.50489 7.74704 -12.7702      Si 7.74586 -2.50925 7.74891
Si -12.7655 12.8854 2.62578        Si -2.50736 12.8784 -7.63801      Si 7.7495 2.62271 12.8819
Si -7.63826 7.75079 2.62543        Si 2.62143 7.74725 -7.63649       Si 12.8749 -2.5071 12.8796
Si -7.63765 12.8831 -2.50746       Si 2.62317 12.8807 -12.7667       Si 12.8807 2.62425 7.75257
```

```
Si 7.75434 7.75079 -12.7678
Si 7.75232 12.8807 -7.63627
Si 12.8854 7.75196 -7.64036
Si 12.8843 12.8867 -12.7691
Si 7.7535 7.74693 -2.50842
Si 7.7528 12.8807 2.61944
Si 12.8854 7.75503 2.62496
Si 12.8831 12.8807 -2.50864
Si 7.75398 7.75162 7.75221
Si 7.75516 12.8831 12.8854
Si 12.8867 7.75468 12.8854
Si 12.8867 12.8843 7.75444
Si -15.335 -15.3327 -15.3374
Si -15.341 -10.203 -10.2078
Si -10.2024 -15.3363 -10.2079
Si -10.2063 -10.2022 -15.3339
Si -15.3374 -15.3363 -5.07645
Si -15.3398 -10.2055 0.05141
Si -10.2039 -15.335 0.0567432
Si -10.2084 -10.2072 -5.07753
Si -15.3327 -15.3339 5.18852
Si -15.3398 -10.202 10.3172
Si -10.1992 -15.335 10.3195
Si -10.2088 -10.2057 5.18698
Si -15.3421 -5.07116 -15.3363
Si -15.3363 0.0582367 -10.2053
Si -10.2082 -5.07292 -10.2021
Si -10.2059 0.0550062 -15.335
Si -15.341 -5.07246 -5.07645
Si -15.3398 0.0578735 0.0560437
Si -10.2139 -5.07753 0.0554214
Si -10.2091 0.0586767 -5.07069
Si -15.3445 -5.07503 5.18395
Si -15.341 0.0563363 10.3166
Si -10.2108 -5.07528 10.3181
Si -10.2122 0.0515958 5.18642
Si -15.3327 5.18712 -15.335
Si -15.3327 10.3193 -10.2058
Si -10.2026 5.18864 -10.2043
Si -10.1969 10.3149 -15.3386
Si -15.3339 5.18995 -5.07292
Si -15.3339 10.3202 0.0603264
Si -10.207 5.18852 0.0606947
Si -10.203 10.32 -5.07221
Si -15.335 5.189 5.18969
Si -15.3303 10.3188 10.32
Si -10.2009 5.18145 10.3171
Si -10.1987 10.3179 5.19111
Si -5.06998 -15.3386 -15.3363
Si -5.07433 -10.2055 -10.2031
Si 0.0571807 -15.3386 -10.2021
Si 0.0536703 -10.2079 -15.335
Si -5.07269 -15.3374 -5.07456
Si -5.07891 -10.207 0.0545524
Si 0.0560589 -15.3386 0.0547146
Si 0.0542583 -10.205 -5.07269
Si -5.07069 -15.3374 5.18662
Si -5.07563 -10.2065 10.3185
Si 0.0591976 -15.3398 10.3158
Si 0.0524004 -10.2074 5.18299
Si -5.07799 -5.07503 -15.3339
Si -5.07881 0.0569327 -10.2014
Si 0.0496696 -5.07446 -10.2026
Si 0.0527731 0.0528906 -15.335
Si -5.08127 -5.07517 -5.07116
Si -5.08234 0.0517994 0.0588517
Si 0.0503036 -5.07645 0.0553003
Si 0.0488854 0.0540384 -5.07269
Si -5.08138 -5.07952 5.1857
Si -5.07635 0.0492758 10.3158
Si 0.0511972 -5.07902 10.3169
Si 0.0524732 0.0485417 5.1857
Si -5.06939 5.18145 -15.3374
Si -5.0708 10.3138 -10.2045
Si 0.0556836 5.18313 -10.2021
Si 0.0607477 10.3139 -15.335
Si -5.07717 5.18616 -5.06951
Si -5.07303 10.3159 0.0589754
Si 0.0536716 5.18017 0.0568822
Si 0.0565656 10.3126 -5.07151
Si -5.07386 5.17935 5.18642
Si -5.06445 10.3117 10.3153
Si 0.0593669 5.18029 10.3157
Si 0.0602006 10.3114 5.18395
Si 5.18662 -15.3374 -15.3339
Si 5.18145 -10.2041 -10.2044
Si 10.3154 -15.335 -10.2038
Si 10.3101 -10.2021 -15.3386
Si 5.18616 -15.3386 -5.07256
Si 5.18452 -10.205 0.0526967
Si 10.3181 -15.3374 0.0539985
Si 10.3129 -10.2021 -5.07785
Si 5.1858 -15.3386 5.18217
Si 5.18145 -10.2082 10.3125
Si 10.3173 -15.335 10.3166
Si 10.3147 -10.2074 5.18195
Si 5.17899 -5.07538 -15.3374
Si 5.18288 0.0555978 -10.2063
Si 10.3122 -5.07126 -10.2083
Si 10.3134 0.0576828 -15.3386
Si 5.18145 -5.07105 -5.07656
Si 5.18277 0.0540609 0.0547747
Si 10.3119 -5.0728 0.0527944
Si 10.3164 0.0579649 -5.0761
Si 5.18171 -5.07727 5.18345
Si 5.18229 0.0540031 10.3164
Si 10.3088 -5.07374 10.313
Si 10.313 0.0576971 5.18544
Si 5.18676 5.18606 -15.335
Si 5.18794 10.3143 -10.2011
Si 10.3195 5.18534 -10.2054
Si 10.3205 10.3187 -15.3327
Si 5.1858 5.18145 -5.07338
Si 5.18794 10.3127 0.0553346
Si 10.3186 5.18676 0.0575924
Si 10.319 10.314 -5.07303
Si 5.18688 5.18299 5.18559
Si 5.19017 10.3157 10.3177
Si 10.3182 5.18911 10.3187
Si    10.3196    10.319    5.18841
```